\documentclass[12pt]{article}
\newfont{\feff}{cmti10}
\begin{document}

\title{Anomalous Scaling of Structure Functions and Dynamic
Constraints on Turbulence Simulations}
\author{Victor Yakhot$^1$ and Katepalli R. Sreenivasan$^2$\\
$^1$Department of Aerospace and Mechanical Engineering\\ Boston
University, Boston 02215\\
$^2$International Center for Theoretical Physics, Trieste, Italy}

\maketitle \begin{abstract} \noindent The connection between
anomalous scaling of structure functions (intermittency) and
numerical methods for turbulence simulations is discussed. It is
argued that the computational work for direct numerical
simulations (DNS) of fully developed turbulence increases as
$Re^{4}$, and not as $Re^{3}$ expected from Kolmogorov's theory,
where $Re$ is a large-scale Reynolds number. Various relations for
the moments of acceleration and velocity derivatives are derived.
An infinite set of exact constraints on dynamically consistent
subgrid models for Large Eddy Simulations (LES) is derived from
the Navier-Stokes equations, and some problems of principle
associated with existing LES models are highlighted.

\end{abstract}
\newpage
\section{Background}

The theory of turbulence and the development of calculation
methods for high-Reynolds-number flows became an active research
topic around the beginning of the twentieth century. This effort
yielded many important results of general interest in statistical
physics. For instance, Kolmogorov's work [1]-[3] on turbulence
theory formulated the scaling ideas for the first time, and
Kraichnan [4] proposed the mode coupling approach. However, the
``turbulence problem", lacking a small parameter characterizing
the strong nonlinear interactions, has turned out to be remarkably
difficult---and it remains so today.

The revolutionary realization of Osborne Reynolds that turbulence
theory is a subject of statistical hydrodynamics rather than
classical hydrodynamics, led almost hundred years ago to various
elegant and useful phenomenological models based on ideas of
kinetic theory (Prandtl [5], Richardson [6], Kolmogorov [3]),
which strongly impacted the engineering profession. These
heuristic semi-empirical models, based on low-order closures of
various perturbation expansions, had a somewhat limited range of
success and needed adjustable parameters, often varying from flow
to flow. Nevertheless, the role of these models was---and still
is---so immense that one can hardly imagine processes in
mechanical and chemical engineering, aerodynamics and meteorology
which do not have their input.

With the advent of powerful computers, the possibility of accurate
numerical simulations, directly based on the Navier-Stokes
equations, became a reality. Since the introduction of spectral
methods in the end of sixties [7]-[8], direct numerical
simulations (DNS) have become a new tool to attack the
``turbulence problem". A strategic goal of the DNS has been to
complement expensive and complicated physical experiments, and
their dream is to dispense with them altogether.

The computational power required for DNS is estimated on the basis
of Kolmogorov's phenomenology that describes turbulent
fluctuations filling the interval of wavenumbers $1/L \ll k \ll
1/\eta_{K}$, where $L$ and $\eta_{K} = LRe^{-\frac{3}{4}}$ are the
integral and dissipation scales, respectively, and $Re =
u_{rms}L/\nu$ is the Reynolds number based on $L$ and the
root-mean-square velocity $u_{rms}$. If we assume that the
velocity fluctuations on scales $r<<\eta_K$ are highly damped and
cannot contribute to the inertial range dynamics, the effective
number of degrees of freedom [9] is then $(L/\eta_K)^3 =
Re^{9/4}$. This is the minimum number of grid points required in
DNS for a cubic box of linear dimension $L$. The required number
of time steps in the computation is usually proportional to the
spatial grid points, so the total computational work increases as
$Re^{3}$. This means that a mere doubling of the Reynolds number
requires almost an order of magnitude increase of computational
work.

The accuracy of numerical methods is traditionally estimated as
follows. The dissipation contribution to the equation for
turbulent kinetic energy is given by
$${\cal E}=-\nu \overline{{\bf u}\cdot \frac{\partial^{2} {\bf u}}{\partial
x_{i}^{2}}} = -\nu \ lim_{r\rightarrow \eta} \frac{\partial
^{2}}{\partial r^{2} }\overline{u_{i}(x)u_{i}(x+r)}= \nu \
lim_{r\rightarrow \eta} \frac{1}{2}\frac{\partial ^{2}}{\partial
r^{2} }S_{2,0}(r)\propto \nu {\cal
E}^{\frac{2}{3}}\eta^{\xi_{2}-2},$$ where the order of magnitude
estimate in the last step comes from Kolmogorov's phenomenology.
For this case, $\xi_{2}=2/3$ and we have $\eta_{K} =
(\frac{\nu^{3}}{{\cal E}})^{\frac{1}{4}}$. We then have the
familiar estimate $\eta_{K}\approx LRe^{-\frac{3}{4}}$, mentioned
earlier. Thus, to accurately describe the flow, one has to simply
account for fluctuations on the scales $r \geq \eta_{K}$ by
choosing the computational mesh size to be
\begin{equation}
\Delta=a\eta_{K}\approx aLRe^{-\frac{3}{4}},
\end{equation}
where $a=const=O(1)$. On this mesh, the velocity derivative is
defined as
\begin{equation}
\frac{u(x+\Delta)-u(x)}{\Delta}=\frac{\partial u(x)}{\partial
x}+\sum_{n=2}\frac{1}{n!}\frac{\partial^{n}u(x)} {\partial
x^{n}}\Delta^{n-1}.
\end{equation}
Now, in Kolmogorov's turbulence, $(\partial_{x}
u)_{rms}=\sqrt{\overline{(\partial_{x} u)^{2}}}\approx
(\frac{{\cal E}Re}{u_{rms}L})^{\frac{1}{2}}=O(Re^{\frac{1}{2}})$,
and, since $\frac{\partial^{n} u(x)}{\partial x^{n}}\approx
\partial_{x} u(x)/\eta_{K}^{n-1}$, using the mesh size $\Delta$
from the relation (1), we arrive at the estimate
\begin{equation}
\frac{1}{n!}(\frac{\partial^{n} u(x)}{\partial
x^{n}})_{rms}\Delta^{n-1}\approx \frac{1}{n!}(\partial_{x}
u)_{rms}(\frac{\Delta}{\eta_{K}})^{n-1}\approx
\frac{a^{n-1}}{n!}Re^{\frac{1}{2}}.
\end{equation}
The relation (3) is essentially the basis for all numerical finite
difference schemes used for the DNS of turbulence [9]. Indeed, we
see that if $a<1$, the first-order finite difference accurately
represents the velocity derivatives.

In spectral simulations of isotropic and homogeneous turbulence,
one prescribes a suitable number of the Fourier modes to represent
the velocity field. Usually, this number is chosen on the basis of
the magnitude of the expected Kolmogorov scale $\eta_{K}$ or the
largest wavenumber $k_{max} = 2\pi/\eta_K$. In the
state-of-the-art simulations [10],[11], the cut-off is usually
chosen such that $k_{max} = \sqrt 2N/3$ on a grid of size $N^3$.

In summary, the principal elements of Kolmogorov's phenomenology
which have enabled these traditional estimates are the following:
(a) the scaling exponents of the structure functions $S_{n,0}
\propto r^{\xi_n}$ are given by the Kolmogorov values
$\xi_{n}=n/3$; (b) the mean dissipation rate ${\cal E}=\nu
\overline{(\partial_{i}u_{j})^{2}}$ is constant and $O(1)$, as are
the moments of the dissipation rate $\overline{{\cal E}^{n}}$ for
all $n$; if the latter were not the case, one can define different
Kolmogorov scales on the basis of different moments of $\cal E$;
and (c) the ``skewness" factors $\overline{(\partial_{x}
u)^{n}}/\overline{(\partial_{x} u)^{2}}^{\frac{n}{2}}=O(1)$,
independent of the Reynolds number; for, if this were not so, one
can again define different Kolmogorov scales through odd moments
of different order.

The main point of the present paper is that there is a need to
reexamine the traditional estimates in the light of modern
developments in turbulent theory and experiment. We concentrate on
isotropic and homogeneous turbulence but expect that the
considerations hold for more general flows as well.

\section{Results for Intermittent Turbulence}
We are interested in the Navier-Stokes dynamics of incompressible
fluids. In 1941, Kolomogorov derived the few exact relation of
turbulence theory, presented here for an arbitrary space
dimensionality $d$, as
$$\frac{1}{r^{d+1}}\frac{\partial}{\partial
r}r^{d+1}S_{3,0}=(-1)^{d}\frac{12}{d}{\cal E},$$ giving
$S_{3,0}=-\frac{12}{d(d+2)}{\cal E}r$ and $S_{3,0}/S_{1,2}=3.$ A
dimensional generalization of this result, without however the
analytical support, yields the Kolomogorov's (normal) scaling
$\xi_{n}=n/3$. Recently [12],[13], some additional exact
consequences of the Navier-Stokes equations have been derived. In
combination with recent experimental results, we consider their
consequences for intermittent turbulence.

{\bf a. Dissipation scale as a random field} We consider the
moments of velocity difference (also called structure functions).
Choosing the displacement vector ${\bf r}$ parallel to the
``$x$-axis", we can define the structure functions
$S_{n,m}(r)=\overline{(u({\bf x}+r{\bf i})-u({\bf x}))^{n}(v({\bf
x}+r{\bf i})-v({\bf x}))^{n}}\equiv
\overline{(\delta_{r}u)^{m}(\delta_{r} v)^{n}}$, where $u$ and $v$
are the components of velocity vector parallel and normal the
$x$-axis, respectively. In the inertial range the velocity
structure functions are $Re$-independent; that is, if the
displacement $r$ belongs to the interval $\eta \ll r \ll L$, then
$S_{n,m}(r)$ do not involve any information about the dissipation
scale.

Modern experiments have revealed that Kolmogorov's result
$\xi_{n}=n/3$ is almost certainly incorrect and that $\xi_{n}$ is
a concave function of $n$---or the ratio $\xi_{n}/n$ is a
decreasing function of the moment number $n$. (See for example
Refs.\ [14] for reviews and Ref.\ [15] for the most recent data.)
Further, the form of structure functions is given by
$S_{2n}(r)=\overline{(u(x+r)-u(x))^{2n}}\approx (2n-1)!!(\epsilon
L)^{\frac{2n}{3}}(\frac{r}{L})^{\xi_{2n}}$. The factor $(2n-1)!!$,
ensuring Gaussian statistics at the integral scale $L$, is a
subject of a forthcoming paper, but it suffices here to say here
that it has been recently verified in experiments and numerical
simulations [16]. On the other hand, in the limit $r\rightarrow
0$, the analytic structure function is approximately equal to
$S_{2n}(r)\approx \overline{(\partial_{x} u(0))^{2n}}r^{2n}$.
Combining the two, we can define a natural dissipation scale of
the $2n^{th}$-order structure function [17]-[18] as
\begin{equation}
\eta_{2n}=(\overline{(\partial_{x}
u)^{2n}})^{\frac{1}{\xi_{2n}-2n}}
((2n-1)!!\epsilon^{\frac{2n}{3}}L^{\frac{2n}{3}-\xi_{n}})^{\frac{1}{2n-\xi_{2n}}}.
\end{equation}
According to (4), the dissipation scales, which are expressed in
terms of the moments of velocity derivatives, define a random
field $\eta$. By a random field we mean here that the value of the
length scale $\eta$ depends on the order of the moment considered.
It will be shown below that (4) is an approximation to a more
accurate representation. Similar ideas were proposed earlier in
Refs.\ [19]-[21] within the framework of multifractal theories.
Writing $i_{2n}=[(2n-1)!!]^{\frac{1}{2n-\xi_{2n}}}$, and using the
Stirling formula $(n\gg1)$, one obtains $i_{2n}\approx
(\frac{n}{2e})^{\frac{3}{4}}$ for $\xi_{n}=n/3$. This means that
the effect of the factor $(2n-1)!!$ can be safely neglected. For
anomalous exponents $\xi_{n}<n/3$, this factor is even closer to
unity and does not modify the conclusions obtained below.

{\bf b.\ Dissipation anomaly} If the velocity field is
differentiable, we obtain $S_{3}(r)\propto r^{3}$ and
$\partial_{r}S_{3}(r)\rightarrow 0$ in contradiction with the
Kolmogorov relation. This implies that the velocity field is
singular in the limit of $\nu\rightarrow 0$ and $r\rightarrow 0$
(in that order), leading to the so-called dissipation anomaly.
Here we first reproduce some details of Polyakov's derivation [22]
of the dissipation anomaly for turbulence governed by Burgers
equation and then outline similar procedure for the Navier-Stokes
equations. Consider the one-dimensional Burgers equation
\begin{equation}
\frac{\partial u}{\partial t} +u\frac{\partial u}{\partial
x}=\nu\frac{\partial^{2} u}{\partial x^{2}},
\end{equation}
for which the energy balance reads as
$$\frac{1}{2}\frac{\partial u^{2}}{\partial
t}+\frac{1}{3}\frac{\partial}{\partial x}u^{3}=\nu
u(x)\frac{\partial ^{2} u}{\partial x^{2}}.$$ Introducing $x_{\pm
}=x\pm\frac{y}{2}$, so that, $\frac{1}{2}\frac{\partial}{\partial
x_{\pm}}=\pm\frac{\partial}{\partial y}$, we can represent the
energy balance  equation as
\begin{equation}
lim_{y\rightarrow 0}[\frac{\partial u(x_{+})u(x_{-})}{\partial t}+
\frac{1}{2}\frac{\partial}{\partial x_{+}}u(x_{+})^{2}u(x_{-})+
\frac{1}{2}\frac{\partial}{\partial
x_{-}}u(x_{-})^{2}u(x_{+})=\nu(\frac{\partial^{2}}{\partial
x_{+}^{2}}+\frac{\partial^{2}}{\partial
x_{-}^{2}})u(x_{+})u(x_{-}))].
\end{equation}
We also have the identities:
\begin{equation}
\frac{\partial}{\partial
y}(u(x_{+})-u(x_{-}))^{3}=\frac{1}{2}[\frac{\partial
u(x_{+})^{3}}{\partial x_{+}}+\frac{\partial
u(x_{-})^{3}}{\partial x_{-}}]-\frac{3}{2}[\frac{\partial
u(x_{+})^{2}u(x_{-})}{\partial x_{+}}+\frac{\partial
u(x_{-})^{2}u(x_{+})}{\partial x_{-}}],
\end{equation}
and
\begin{equation}
\nu [u(x_{+})\frac{\partial^{2} u(x_{-})}{\partial
x_{-}^{2}}+u(x_{-})\frac{\partial^{2} u(x_{+})}{\partial
x_{+}^{2}}]= \nu[(u(x_{+})-u(x_{-}))\frac{\partial^{2}}{\partial
y^{2}}(u(x_{+})-u(x_{-}))]+D,
\end{equation}
where $D$, the dissipation contribution to the energy balance, is
given by
\begin{equation}
D=\nu[u(x_{+})\frac{\partial^{2}}{\partial
x_{+}^{2}}u(x_{+})+u(x_{-})\frac{\partial^{2}}{\partial
x_{-}^{2}}u(x_{-})]. \end{equation} Substituting these identities
into the equation (6) and taking account of the fact that
$lim_{y\rightarrow 0} \frac{\partial u(x_{\pm})^{3}}{\partial
x_{\pm}}= \frac{\partial u(x)^{3}}{\partial x}$, so that in the
limit $y\rightarrow 0$ all non-singular terms disappear by virtue
of the energy equation (5), we are left with the balance between
the singular (anomalous) contributions
\begin{equation}
\lim_{y\rightarrow 0}\frac{1}{6}\frac{\partial}{\partial
y}(u(x_{+})-u(x_{-}))^{3}=
\nu[(u(x_{+})-u(x_{-}))\frac{\partial^{2}}{\partial
y^{2}}(u(x_{+})-u(x_{-}))].
\end{equation}
This is Polyakov's expression for the dissipation anomaly derived
for the Burgers equation [22]. Averaging (10) gives the exact
relation $\overline{(\delta_{y} u)^{3}}=-12{\cal E}y$ where the
dissipation rate ${\cal E}=\nu\overline{(\frac{\partial
u}{\partial x})^{2}}$.

We are interested in the Navier-Stokes dynamics of incompressible
fluids, for which the energy balance equation (with the density
$\rho=1$) is written as
$$\frac{1}{2}\frac{\partial
u^{2}}{\partial t}+\frac{1}{2}{\bf u}\cdot \nabla u^{2} =-\nabla
p\cdot {\bf u}+ \nu {\bf u}\cdot \frac{\partial ^{2} {\bf
u}}{\partial x_{i}^{2}},$$ and that for the scalar product ${\bf
u(x+\frac{y}{2})\cdot u(x-\frac{y}{2})\equiv u(+)\cdot u(-)}$ can
be written as
\begin{eqnarray}
\frac{\partial {\bf u(+)\cdot u(-)}}{\partial t}+{\bf
u(+)\cdot\frac{\partial}{\partial  x_{+}}u(+)\cdot u(-)}+{\bf
u(-)\cdot\frac{\partial}{\partial x_{-}}u(-)\cdot u(+)}=\nonumber
\\-\frac{\partial p(+)}{\partial x_{+,i}}u_{i}(-)-\frac{\partial
p(-)}{\partial x_{-,i}}u_{i}(+)+ \nu[{\bf u}(-)\cdot
\frac{\partial^{2}}{\partial x_{+,j}^{2}}{\bf u}(+)+{\bf u}(+)\cdot
\frac{\partial^{2}}{\partial x_{-,j}^{2}}{\bf u}(-)].
\end{eqnarray}
It is clear that in the limit $y\rightarrow 0$, for which ${\bf
x}_{\pm}\rightarrow {\bf x}$, this equation gives the energy
balance. Following Polyakov's procedure outlined above, let us
consider the two identities:
\begin{eqnarray}
\frac{\partial}{\partial y_{i}}(u_{i}(+)-u_{i}(-))(u_{j}(+)-u_{j}(-))^{2}=\nonumber \\
\frac{1}{2}\frac{\partial}{\partial
x_{+,i}}u_{i}(+)u_{j}^{2}(+)+\frac{1}{2}\frac{\partial}{\partial
x_{+,i}}u_{i}(+)u_{j}^{2}(-)-\frac{\partial}{\partial
x_{+,i}}u_{i}(+)u_{j}(+)u_{j}(-)+\nonumber \\
\frac{1}{2}\frac{\partial}{\partial
x_{-,i}}u_{i}(-)u_{j}^{2}(-)+\frac{1}{2}\frac{\partial}{\partial
x_{-,i}}u_{i}(-)u_{j}^{2}(+)-\frac{\partial}{\partial
x_{-,i}}u_{i}(+)u_{j}(-)u_{j}(+)
\end{eqnarray}
and
\begin{eqnarray}
u_{i}(+)\frac{\partial^{2}}{\partial x_{-,j}^{2}}
u_{i}(-)+u_{i}(-)\frac{\partial ^{2}}{\partial
x_{+,j}^{2}}u_{i}(+)=\nonumber
\\
-4(u_{i}(+)-u_{i}(-))\frac{\partial^{2}}{\partial
y_{j}^{2}}(u_{i}(+)-u_{i}(-))+u_{i}(+)\frac{\partial^{2}}{\partial
x_{+,j}^{2}}u_{i}(+)+ \nonumber  \\
u_{i}(-)\frac{\partial^{2}}{\partial x_{-,j}^{2}}u_{i}(-).
\end{eqnarray}
Similar identities for the pressure terms can be written easily.
Substituting them into (11) and, as in the case of Burgers
equation considered above, accounting for the energy balance, one
has
\begin{eqnarray}
lim_{y\rightarrow 0}[-\frac{\partial}{\partial
y_{i}}(u_{i}(+)-u_{i}(-))(u_{j}(+)-u_{j}(-))^{2}+
\frac{1}{2}(\frac{\partial}{\partial
x_{+,i}}u_{i}(+)u_{j}(-)^{2}+\frac{\partial}{\partial
x_{-,i}}u_{i}(-)u_{j}(+)^{2}) = \nonumber \\
-4\nu(u_{i}(+)-u_{i}(-))\frac{\partial^{2}}{\partial
y_{j}^{2}}(u_{i}(+)-u_{i}(-))+(\frac{\partial p(+)}{\partial {\bf
x_{+}}}-\frac{\partial p(-)}{\partial {\bf x_{-}}})\cdot ({\bf
u(+)-u(-)})].
\end{eqnarray}
This equation can be written in a compact form as
$$
lim_{y\rightarrow 0}[- \frac{\partial}{\partial y_{i}}\delta
u_{i}|{\bf \delta_{y} u}|^{2}+\frac{1}{2}(\frac{\partial}{\partial
x_{+,i}}u_{i}(+)u_{j}(-)^{2}+\frac{\partial}{\partial
x_{-,i}}u_{i}(-)u_{j}(+)^{2})=-2{\bf \delta_{y} u \cdot \delta_{y}
a}],
$$
where ${\bf a}=-\nabla p+\nu \nabla^{2}{\bf u}$ is the Lagrangian
acceleration. The equation (14) is exact. Choosing the
displacement vector along one of the coordinate axes and averaging
(14), one obtains
$$\frac{\partial }{\partial y} \overline{\delta u |{\bf \delta
u}|^{2}}=8\overline{\delta u_{i}\frac{\partial^{2}}{\partial
y^{2}}\delta
u_{i}}=2\overline{(\delta_{y}u_{i})\partial_{x}^{2}(\delta_{y}u_{i}})=-\frac{4}{3}{\cal
E},$$ where $\delta_{y} u={\bf \delta_{y} u} \cdot {{\bf y}/y}$.
The pressure terms in and the second contribution to the left side
of (14) disappeared by the averaging procedure. In general, we can
choose a sphere of radius $y<<R\rightarrow 0$ around a point ${\bf
x}$ and average (14) over this sphere. This causes the all
scalar-velocity contributions to (14) disappear and the resulting
equation can be perceived as a local form of the $4/3$ Kolmogorov
law. This fact has been realized before. Introducing the angular
averaging, Robert and Duchon [23] and Eyink [24] locally expressed
the relation (14) in terms of longitudinal and transverse velocity
differences. We are interested in the order of magnitude estimates
(see below), and restrict ourselves to (14).

{\bf c.\ Relations between the moments} In the isotropic and
homogeneous turbulence, the Navier-Stokes equations lead to the
following exact relations for structure functions. They were
derived in [12] and [13] and experimentally investigated in some
detail in Ref.\ [25]; see also Ref.\ [26]. The relations for
different values of $n$ are
\begin{equation}
\frac{\partial S_{2n,0}}{\partial
r}+\frac{d-1}{r}S_{2n,0}=\frac{(2n-1)(d-1)}{r}S_{2n-2,2}+
(2n-1)\overline{\delta_{r}a_{x}(x)(\delta_{r}u)^{2n-2}}.
\end{equation}
Similar equations for all structure functions $S_{n,m}$ can easily
be obtained from the equation for generating functions derived in
[12].

{\bf d.\ The closure problem} Equation (15), which includes both
velocity and Lagrangian acceleration increments, is not closed and
cannot be solved unless the relation between acceleration and
velocity differences is established. It has been proposed in Ref.\
[17] that the local expression (14) written for the displacement
magnitudes corresponding to the bottom of inertial range, i.e. in
the limit $y\rightarrow \eta \rightarrow 0$ can be used as a
closure. At the present time, this can be done only approximately.
Since at the values of displacement $y\ll\eta\rightarrow 0$, the
difference $\delta_{y}u\approx \frac{\partial u(0)}{\partial x}
y$, we can modify the $lim$ operation in (14) as
\begin{equation}
\lim_{y\rightarrow 0}\approx \lim_{y\rightarrow\eta\rightarrow 0},
\end{equation}
leading to the order-of-magnitude estimate
\begin{equation}
lim_{y\rightarrow \eta} A\frac{\partial (\delta_{y}
u)^{3}}{\partial y} +B\frac{\partial}{\partial y} \delta_{y}
u(\delta _{y} v)^{2}\propto
\nu\delta_{y}u\frac{\partial^{2}}{\partial y^{2}} \delta_{y}
u-\frac{\partial \delta_{y} p(x)}{\partial y}\delta_{y} u\approx
\delta_{\eta} u\delta_{\eta} a_{x},
\end{equation}
where $A$ and $B$ are undetermined constants. On extrapolating to
the dissipation scale $\eta$ where all terms in the right side of
(18) are of the same order, we derive the estimate [17] as
\begin{equation}
\nu\approx \eta\delta_{\eta} u\equiv\eta(u(x+\eta)-u(x)).
\end{equation}
The relation (18) tells us that each velocity fluctuation
$\delta_{\eta} u$ is dissipated on its `own'  dissipation scale
$\eta$ and the local value of the Reynolds number $Re_{l}=O(1)$.
This allows a simple physical interpretation that the dissipation
processes at all levels $n$ happen on ``quasi-laminar structures"
where the inertial and viscous terms are of the same order. In
general, the higher the moment order, the more the intense events
contribute, and the smaller the value of the corresponding
dissipation scale.

{\bf e.\ Dissipation scales and moments of derivatives} The theory
gives for the moments of Lagrangian acceleration ${\bf a}=-\nabla
p+\nu \nabla^{2}{\bf u}$ the result that
\begin{equation}
a_{x}\approx  \frac{\delta_{\eta} u}{\tau_{\eta}}\approx
\frac{(\delta_{\eta} u)^{2}}{\eta}\approx
\frac{(\delta_{\eta}u)^{3}}{\nu}=(\delta_{\eta}
u)^{3}\frac{Re}{u_{rms}L},
\end{equation}
where the turn-over time $\tau_{\eta}\approx \eta/\delta_{\eta}u$.

Below we will mainly discuss the equations for even-order
structure functions, for which, if the displacement $r$ is in the
inertial range, the dissipation contribution to the increment of
Lagrangian acceleration is negligibly small [17],[18]. For this
case, we have
\begin{equation}
\frac{\partial S_{2n,0}}{\partial
r}+\frac{d-1}{r}S_{2n,0}=\frac{(2n-1)(d-1)}{r}S_{2n-2,2}-
(2n-1)\overline{\delta_{r} p_{x}(\delta_{r}u)^{2n-2}},
\end{equation}
where $p_{x}=\partial_{x}p(x)$ and $d$ denotes, as before, the
space dimensionality.

The relation (15) is valid for all magnitudes of displacement
$r\ll L$, including $r\rightarrow \eta$. Below, to simplify the
notation, we will omit the subscript $x$ in the $x$-component of
acceleration $a_{x}$. In this limit, treating (19) as
$a=\lim_{r\rightarrow \eta} (\delta_{r} u)^{3}/\nu$ and
substituting it in (15) gives $\frac{S_{2n}(r)}{r}\approx
\frac{S_{2n+1}(r)}{\nu}$. On a scale $r=\eta_{2n}$, writing
$S_{n,0}\propto  A_{n}\eta_{n}^{\xi_{n}}$, equation (15) gives
\begin{equation}
\eta_{n}\propto LRe^{\frac{1}{\xi_{n}-\xi_{n+1}-1}}.
\end{equation}

For Kolmogorov turbulence with $\xi_{n}=n/3$ the formula (21)
reads, as expected, as $\eta_{n}\equiv\eta_{K} =
LRe^{-\frac{3}{4}}$ which is $n$-independent. In intermittent
turbulence, where the exponents can be well-described [12],[17] by
the relation $\xi_{n}\approx 0.383*n/(1+0.05n)$, the relation (21)
defines the Reynolds-number-dependent dissipation scales. As
$n\rightarrow\infty$, $\eta_{n}\rightarrow LRe^{-1}$. Thus, to
resolve all fluctuations including the strongest, the
computational work need to increase as $Re^{4}$, as already noted
in Ref.\ [27]. In general, in the limit $n\rightarrow \infty$, the
relation (21) can be written as
$$\eta_{n} \approx LRe^{-\frac{1}{\frac{d \xi_{n}}{dn}+1}},$$
so one may get a somewhat different estimate for the computational
work than $Re^4$, but the principal conclusion is inescapable that
intermittency makes DNS more expensive than previously thought.

Using the relations (18), (20) and (21), obtained by balancing
various terms in the exact dynamic equations (14), (15), we can
develop the multi-scaling algebra. For example,
\begin{equation}
\overline{a^{2n}}\approx
(\frac{Re}{u_{rms}L})^{2n}S_{6n}(\eta_{6n})\propto
(\frac{Re}{u_{rms}L})^{2n}\eta_{6n}^{\xi_{6n}}\approx
(\frac{u_{rms}^{2}}{L})^{2n}Re^{a_{2n}},
\end{equation}
with $a_{2n}=2n+\frac{\xi_{6n}}{\xi_{6n}-\xi_{6n+1}-1}$. With
$\xi_{6}=2$ and $\xi_{7}=7/3$, we recover Yaglom's result [28]
$\overline{a^{2}}\approx
\frac{u_{rms}^{\frac{9}{2}}}{\sqrt{\nu}}$. The intermittency
corrections are readily found from (22). Recent experiments by
Reynolds et al.\ [29] have lent strong support to this result. The
formula (22) shows that the second moment of Lagrangian
acceleration is expressed in terms of the sixth-order structure
function evaluated on its dissipation scale $\eta_{6}$. To extract
information about the fourth moment $\overline{a^{4}}$, we should
have accurate data on $S_{12}(\eta_{12})$ which is very difficult
to obtain in high-Reynolds-number flows.

The moments of velocity derivatives are evaluated easily. In
accordance with (18), we have
\begin{equation}
\overline{(\partial_{x} u)^{2n}}\approx
\overline{(\frac{\delta_{\eta} u}{\eta})^{2n}}\approx
\overline{(\frac{(\delta_{\eta} u)^{2}}{\nu})^{2n}}\approx
Re^{d_{2n}},
\end{equation}
where $d_{2n}=2n+\frac{\xi_{4n}}{\xi_{4n}-\xi_{4n+1}-1}$.

It is important to stress that the first equality in (23) involves
the averaging over two random fields $u$ and $\eta$. To perform
this averaging, we have to either know the joint probability
$p(u,\eta,r)$ or use the functional relation between the fields
given by (18). This leads to the second equation in (23) and the
final result. Since $\overline{(\partial_{x}u)^{2}}\propto Re$,
the relation (3) leads  to a new relation between exponents
$$2\xi_{4}=\xi_{5}+1$$ which agrees extremely well with experimental
data. The relation (23) differs from proposals reviewed in Ref.\
[14].

{\bf f.\ The role of the fluctuations of the dissipation scale}
Let us reexamine the relation (4). In the limit $r\rightarrow 0$,
the velocity field is analytic and can be expanded by Taylor
series so that $\frac{\partial u}{\partial x}\approx
\delta_{r}u/r$. This gives $\overline{(\frac{\partial u}{\partial
x}r)^{2n}}\approx S_{2n}(r)$. When $r\rightarrow \eta\rightarrow
0$, we have to evaluate the mean of the ratio
$\overline{(\delta_{\eta} u/\eta)^{2n}}$ which is not a trivial
task, since we are dealing here with the ratio of two random
fields---unless the relation (18), which expresses the dissipation
scale in terms of velocity field, is used. If, however, we {\it
incorrectly} assume that the dissipation scale fluctuations are
independent of those of the velocity field and neglect the step
leading to the last equations in the right hand side of (23), it
is possible to write the moments of velocity derivative as
\begin{equation}
\overline{(\partial_{x}u)^{2n}}\approx
\overline{(\frac{\delta_{\eta} u}{\eta})^{2n}}\approx
S_{2n}(\eta_{2n})/\eta_{2n}^{2n}\propto Re^{p_{2n}},
\end{equation}
where $p_{2n}=\frac{\xi_{2n}-2n}{\xi_{2n}-\xi_{2n+ 1}-1}$.
Equating expressions (23) and (24), we have
\begin{equation}
\frac{\xi_{2n}-2n}{\xi_{2n}-\xi_{2n+1}-1} =
2n+\frac{\xi_{4n}}{\xi_{4n}-\xi_{4n+1}-1},
\end{equation}
subject to the constraints $\xi_{0}=0$ and $\xi_{3}=1$. The only
solution to (25) is $\xi_{n}=n/3$. Since equation (25) is based on
the first equality (23), which in general is incorrect, we can
conclude that the source of anomalous scaling in hydrodynamic
turbulence is the fluctuation of the dissipation scale field
$\eta$, which itself is strongly correlated the velocity field
fluctuations via expression (18). This does not preclude a
different situation from arising in other forms of turbulence,
e.g., scalar turbulence generated by white-noise forcing [30].

It follows that $\overline{(\frac{\partial u}{\partial
x})^{2}}=\lim_{r\rightarrow \eta_{2}} \overline{\frac{\partial
u(x)}{\partial x}\frac{\partial u(x')}{\partial
x'}}=-\lim_{r\rightarrow \eta_{2}} \frac{\partial ^{2}}{\partial
r^{2}} \overline{u(x)u(x')}\propto
(2-\xi_{2})\eta_{2}^{\xi_{2}-2}.$ The higher-order derivatives are
evaluated in a similar way to yield
\begin{equation}
(\frac{\partial^{n} u}{\partial x^{n}})_{rms}=lim_{r\rightarrow
\eta_{2}} \sqrt{\frac{\partial^{2n}}{\partial
r^{2n}}S_{2}(r)}\approx \eta_{2}^{\frac{\xi_{2}-2n}{2}}\approx
Re^{\frac{\xi_{2}-2n}{2(\xi_{2}-2)}}=Re^{\frac{1}{2}}Re^{\frac{n-1}{\xi_{2}-2}}.
\end{equation}

\section{Implications for Numerical Methods}
According to experimental data (see Refs.\ [25,15] for recent
results), the exponent $\xi_{2}\approx 0.70-0.71 >2/3$ and as
$n\rightarrow \infty$, the terms in the expansion (2) for
simulating the ``typical" velocity derivatives can be estimated
via
\begin{equation}
(\frac{\partial^{n} u}{\partial x^{n}})_{rms}\Delta^{n-1}\propto
Re^{\frac{1}{2}}Re^{\gamma (n-1)},
\end{equation}
with $\gamma = (-\frac{3}{4}-\frac{1}{\xi_{2}-2}))>0$. For
$\xi_{2}\approx 0.71$, we find $\gamma \approx 0.025$. The
accuracy of the numerical method in calculating the most intense
velocity fluctuations can be estimated if, in the limit
$n\rightarrow \infty$, the expression
\begin{equation}
\overline{(\frac{\partial u}{\partial
x})^{2n}}^{\frac{1}{2n}}(\frac{\Delta}{\eta_{2n}})^{n-1}\propto
Re^{\frac{1}{2}}Re^{\frac{n+1}{4}}
\end{equation}
is used instead of $(\partial_{x} u)_{rms}$. In the above
equation, the mesh size $\Delta$ is defined by (1) and the
expressions (23) for the moments of velocity derivative have been
used. We see that when the Reynolds number is large, the
high-order derivatives in the expression (2) dominate. This means
that the DNS based on the mesh equal to the Kolmogorov scale
becomes quite inaccurate. It is easy to check that accurate
simulations of the largest fluctuations requires the resolution of
the smallest scales which are $O(1/Re)$. This means that the
computational resolution scales as $Re^{3}$ and the computational
work grows as $Re^{4}$.

In Refs.\ [19], it has argued that the intermittent nature of
turbulence makes the size of the attractor smaller than the
conventionally estimated, so the computational power needed
becomes correspondingly smaller than the conventional
estimate---not larger as just claimed. The rationale is roughly
that the ``interesting" parts of the flow occupy small volumes of
space so any reasonable computational effort that focuses on those
volumes is likely to be less expensive. This is also the spirit of
adaptive meshing [31]. Even if the interesting parts of a
turbulent flow are not space-filling, as discussed at length in
Ref.\ [20], we do not yet know how to track them efficiently in
hydrodynamics turbulence. We also do not know if the part of the
flow that contains the less interesting parts can be computed with
greater economy. Nevertheless, it must be said that the present
estimates apply to uniform meshing, which has been the most
successful of the computing schemes until now. It should also be
mentioned that there is a specific suggestion [32] on the most
singular structure in turbulence, which yields $Re^{3.6}$, which
is slightly different from $Re^4$ estimated in this paper.

\section{ Dynamic Constraints on Sub-Grid Models for LES}
If the Reynolds number is large, the computational work involved
in the numerical simulation of a flow is huge. It is interesting
that at about the same time that DNS came into being, the idea of
the Large Eddy Simulations (LES) was proposed by Deardorff [33].
The idea is very simple. Consider the Navier-Stokes equations
\begin{equation}
\partial_{t}{\bf u}+u_{i}\partial_{i}{\bf u}=-
\nabla p +\nu\partial^{2}{\bf u};~~\partial_{i}u_{i}=0.
\end{equation}
We choose the mesh size $\Delta$ and define the so-called
``sub-grid" velocity fluctuations $u^{>}(k)\neq 0$ for $k\geq
\pi/\Delta$. The Fourier-transform of velocity field is defined as
\begin{equation}
u({\bf k})=u^{<}({\bf k})+u^{>}({\bf k}),
\end{equation}
so that
\begin{equation}
u^{>}({\bf x})=\int_{|k|>\frac{2\pi}{\Delta} }e^{i{\bf k}\cdot
{\bf x}}u^{>}({\bf k})d^{3}k;~~~~u^{<}({\bf
x})=\int_{|k|\leq\frac{2\pi}{\Delta} }e^{i{\bf k}\cdot {\bf
x}}u^{<}({\bf k})d^{3}k.
\end{equation}
The goal is to obtain the correct equation for the resolved scales
$u^{<}(k)\neq 0$ in the interval $0\leq k\leq \pi/\Delta$. We
decompose the field and write the equation for only the resolved
scales as
\begin{equation}
\partial_{t}{\bf u^{<}}+u^{<}_{i}\cdot \partial_{i}{\bf u^{<}}={\cal SG}
-\nabla p^{<} +\nu\partial^{2}{\bf u^{<}},
\end{equation}
where, for this particular formulation, the subgrid contribution
is $ {\cal SG}=-u_{i}^{<}\cdot \partial_{i}{\bf u^{>}} -
u_{i}^{>}\cdot \partial_{i}{\bf u^{<}} -u_{i}^{>}\cdot
\partial_{i}{\bf u^{>}}$. The LES equations are considered a
success if the large-scale velocity fields (i.e., for $k\leq
1/\Delta$) given by the Navier-Stokes equations (30) and by a
model (33) are identical or close enough for all Reynolds numbers.

There is, however, one problem. To derive the equation of motion
containing only the resolved fields, one has to express all
contributions to ${\cal SG}$, involving the sub-grid velocity
fluctuations ${\bf u}^{>}$, in terms of ${\bf u}^{<}$, which is
basically equivalent to solution of the proverbial ``turbulence
problem". The model equation (33) is written in a generic form,
but a similar difficulty arises if, instead of the Fourier-space
decomposition introduced above, the filtering or any other kind is
used.

The accurate LES model must satisfy the following dynamic
constraints. The method developed in the Ref.\ [17] can be
literally applied to the Navier-Stokes equations with an arbitrary
right hand side and, defining the coarse-grained structure
functions $S_{n,0}^{<}(r)=\overline{(\delta_{r} u^{<})^{n}}$, we
obtain, from (21), the result
\begin{equation}
\frac{\partial S^{<}_{2n,0}}{\partial
r}+\frac{d-1}{r}S^{<}_{2n,0}=\frac{(2n-1)(d-1)}{r}S^{<}_{2n-2,2}+
(2n-1)\overline{(\delta_{r}({\cal SG}_{x})-\delta_{r}
p^{<}_{x})(\delta_{r}u^{<})^{2n-2}}.
\end{equation}
The large-scale velocity fields obtained from DNS and LES can be
identical $S_{n,0}(r)=S_{n,0}^{<}(r)$ if and only if
\begin{equation}
\overline{(\delta_{r}({\cal SG}_{x})-\delta_{r}
p^{<}_{x})(\delta_{r}u^{<})^{2n-2}}=- \overline{\delta_{r}
p_{x}(\delta_{r}u)^{2n-2}}.
\end{equation}

Similar constraints, coming from the equations for various
structure functions $S_{n,m}$ can be readily obtained. It is
impossible to demand equality of two random fields ${\bf u}$ and
${\bf u}^{<}$ obtained from two different equations. The only
criterion we can impose is that of statistical equality or,
equivalently, constraint on all moments, namely
$S_{n}^{<}(r)=S_{n}(r)$. The relations (34), reflecting this
necessary condition of the LES validity, must be satisfied.

We wish to stress that these constraints are not dissimilar to
$S^{LES}_{n,m} \approx S^{<}_{n,m}$, often implied in the
literature. Here $S_{n,m}^{LES}(r)$ are the structure functions
evaluated from the velocity field obtained from LES. The velocity
increment can be written as $\delta_{r}u=\int
u(k)e^{ikx}(e^{ikr}-1)$, so that
$$S_{2}\propto \int E(k)(1-\cos kr)dk.$$
It is easy to see that if $r<<L$, where $L$ is the integral scale,
and the energy spectrum decreases with $k$ fast enough, the main
contribution to the integral comes from the range where $kr\approx
1$. Thus the structure functions $S_{n,0}(r)$ probe structures on
the scales of the order $r$ and cannot differ strongly from the
one obtained from the filtered field.

Various model considerations, leading to expressions for ${\cal
SG}$, have been suggested in the last forty years. Consider the
example that follows from Kolmogorov's theory. If the role of the
small scale fluctuations in the large-scale dynamics can be
expressed in terms of effective viscosity $\nu_{SG}$, then
$\nu_{SG}\approx (\overline{\epsilon} \Delta^{4})^{\frac{1}{3}}$.
Then, dropping the averaging sign (quite an assumption!) and
substituting a simple estimate coming from the energy balance,
namely, $\epsilon=\nu_{SG}S^{<}_{ij}S^{<}_{ij}\equiv
\nu_{SG}S_{ij}^{2}$, we derive the Smagorinsky formula [34] given
by $\nu_{SG}=\alpha \sqrt{S^{<}_{ij}S^{<}_{ij}}\Delta^{2}$, where
$\alpha=O(1)$. It is important that the resolved rate of strain is
evaluated in terms of velocity differences on the computational
mesh
\begin{equation}
S^{<}_{ij}(x)=\frac{1}{2}(
\frac{u^{<}_{i}(x+\Delta_{j})-u^{<}_{i}(x)}{\Delta_{j}}
+\frac{u^{<}_{j}(x+\Delta_{i})-u^{<}_{j}(x)}{\Delta_{i}}),
\end{equation}
where $i,j=1,2,3$. In this approximation, the Reynolds stress
$\tau_{ij}=-\overline{u_{i}u_{j}}\approx \nu S_{ij}\approx
\nu_{SG}S^{<}_{ij}$. Equation (36) with the model for ${\cal SG}$
defines a closed set of equations which can be used for LES. The
analytically evaluated coefficient from Yakhot and Orszag [35]
gives $\alpha\approx 0.2$, while the so-called dynamic method [36]
gives something different. In all approaches, since the
large-scale fields $\delta_{r}{\bf u}^{<}$ and $\delta_{r}{\bf u}$
are statistically independent upon Reynolds number, the parameter
$\alpha=O(Re^{0})$. Thus, this simple model is
\begin{equation}
{\cal SG}\approx a\Delta^{2}\nabla|S^{<}_{ij}|\nabla u^{<}=O(1).
\end{equation}

Examining the relations (34) and (36), an interesting conclusion
can be reached. If $\Delta\ll r$, one can assume statistical
independence of all velocity differences $\delta_{r} u^{<}$ and
$\delta_{\Delta}u^{<}$. Since ${\cal SG}$ given by (34) and (35)
depends on the velocity differences defined on the mesh size
$\Delta$ as
\begin{equation}
\overline{\delta_{r}{\cal SG}(\delta_{r} u^{<})^{2n-2}}\approx
\overline{\delta_{r}{\cal
SG}}~\overline{(\delta_{r}u^{<})^{2n-2}}=0,
\end{equation}
we see that the Smagorinsky model satisfies the dynamic
constraints, provided the pressure gradient differences in the
filtered and unfiltered fields are close to each other. The
validity of the dynamic Smagorinsky models in the range
$k<<1/\Delta$ has been verified by large eddy simulations (A.
Oberai, private communication 2005). However, as $r\rightarrow
\Delta$, $\delta_{r}{\cal SG}$, $\delta_{r} p_{x}$ and $\delta_{r}
u^{<}$ are strongly correlated and, as a result, the model becomes
invalid. This consideration is applicable to all low-order
closures.

This intrinsic failure of all {\bf existing} LES models at scales
comparable to the computational mesh is well-known. At
sufficiently low Reynolds numbers, LES give accurate results.
However, with increase of $Re$ the quality of the simulations
deteriorates starting from the vicinity of the cut-off,
propagating toward the larger scales. At this point one is forced
to increase the resolution. The reasons for this failure can be
qualitatively understood as follows. Consider LES at a relatively
low Re on a fixed mesh $\Delta/L_{1}=\gamma_{1}$ where $L_{1}$ is
an integral scale of this particular simulation. Now increase the
length scale of the flow $L_{2}>>L_{1}$, thus increasing the
Reynolds number. If, in the first case, the number of the cascade
steps for the energy flux to reach the mesh scale was say $n_{1}$,
that in the second simulation is equal to $n_{2}>>n_{1}$. Since
the intermittency and deviation from the close-to-Gaussian
statistics, experimentally observed at the integral scale, grows
with the number of cascade steps, the contribution from the very
strong velocity fluctuations at the ``dissipation" scale $\Delta$
increases. As a result, the low order models that are successful
in the close to Gaussian situations break down. In another
scenario, let us increase the Reynolds number by increasing the
mean velocity while keeping both the energy injection scale and
the mesh size $\Delta$ constant. In this situation, the top of the
``inertial" range will move into the range of scales which are
larger than $\Delta$, thus again invalidating the LES.

A recent paper by Kang et al.\ [37] has demonstrated that, at the
scales close to those of the mesh size, the probability density
function $p(\delta_{r} u)$ computed from LES was quite close to a
Gaussian while the experimental PDF showed broader tails, typical
of intermittency. This means that the contributions from strong
velocity fluctuations obtained from LES are underpredicted. Since
the intermittent effects becomes stronger with increasing Reynolds
number, we expect this difference to grow, thus invalidating the
LES if the mesh size is also not modified. A very interesting
example is given by the LES of the flow in a simple cavity
reported by Larcheveque et al.\ [38]. It was shown that to
correctly reproduce the experimental data on pressure fluctuations
in a frequency range $100\leq f\leq 2000Hz$, the optimal cut-off
of the large eddy simulations corresponded to $\Delta_{f}=100KHz$.
With decrease of $\Delta_{f}$, the quality of the results rapidly
deteriorated. The present theory explains the failure of LES
schemes with fixed mesh to describe the high Reynolds number flows
as originating from the failure of low-order models in an
all-important range $r\approx \Delta$, this range being
responsible for the energy cascade dissipation. At the present
time, it is not clear how many constraints (35) must be satisfied
to achieve accurate LES, but we believe that the number must grow
with the Reynolds number.

\section{Conclusions}
For many years, intermittency and anomalous scaling in
three-dimensional turbulence were considered major challenges for
theorists. Recent developments of the multifractal theory and its
dynamic formulation led to description of intermittency in terms
of an infinite number of dissipation scales (ultraviolet
cut-offs). It was shown that strong velocity fluctuations are
dissipated on scales that are much smaller than that estimated
from Kolomogorov's theory. In this paper, we have attempted to
make a connection between the theory of anomalous scaling and
numerical methods.

One conclusion that follows from this connection is that to
simulate all fluctuations, including the strongest ones, the
computational demands scale as $Re^{4}$, and not as $Re^{3}$ as
traditionally deduced according to the Kolmogorov theory. To
achieve the full DNS of turbulence, including the strongest
small-scale velocity fluctuations, one has to use resolutions high
enough to produce an analytic interval of structure functions,
where $S_{n}\approx \overline{\partial_{x} u(0))^{n}}r^{n}$.
Analyzing the results of various numerical state-of-the-art DNS,
we have discovered that this criterion is satisfied only for
$n\leq 4$. This is not sufficient to accurately simulate the
velocity derivatives.

A second comment concerns the Large Eddy Simulations. An infinite
number of dynamic constraints on a correct subgrid model has been
derived from the exact relations for structure functions. Due to
the Galilean invariance, the subgrid scales cannot influence the
advective term in the Navier-Stokes equations, provided the
subgrid scale $\Delta/r\rightarrow 0$. However, it is clear from
analyzing the equations of Section 4 that the subgrid model cannot
be reduced to a low-order viscosity expression, but must include
high-order nonlinear contributions that do not vanish at the
scales close to the mesh size.

Thus, while accurate DNS are possible if the resolution
requirements are met and powerful enough computers are available.
However, due to the basic theoretical problems, derivation of an
accurate and theoretically justified subgrid model, valid at very
high Reynolds numbers, remains a major challenge.

It is worth pointing out that we have considered homogeneous and
isotropic turbulence. The situation with wall flows is even more
complex. There, turbulence is mainly produced in the vicinity of
the wall where acceleration and turbulence production are highly
intermittent. Recent DNS by Lee et al.\ [39] have demonstrated
strong intermittency and the Reynolds number dependence of the few
first moments of Lagrangian acceleration near the wall, sharply
peaking at the reduced normalized distance $y_{+}\approx 2.5$. At
present, we do not know how to model this near-wall phenomenon
that is largely responsible for turbulence production.

We wish to conclude on a ``positive" note. The fact that the
structure functions $S_{2n}\approx
(2n-1)!!(\frac{r}{L})^{\xi_{2n}}$ means that the velocity
distribution is close to the Gaussian near $r = L$, and the
intermittency is weak or nonexistent. It follows that simple,
semi-qualitative resummations of the expansions in powers of the
dimensionless rate-of-strain are much less problematic there.
Thus, the derivation of the VLES or time-dependent RANS appears to
have a brighter future.\\

\noindent {\bf References}\\

\noindent 1.\ A.N. Kolmogorov,\  Dokl.\ Akad.\ Nauk SSSR. {\bf
30}, 9 (1941)\\

\noindent 2.\ A.N. Kolmogorov,\ Dokl.\ Akad.\ Nauk SSSR. {\bf
32}, 16 (1941)\\

\noindent 3.\ A.N. Kolmogorov,\ Izv.\ Akad.\ Nauk SSSR.\ Ser.\
Fiz.\
${\bf IV}$ (1-2), 56 (1942)\\

\noindent  4.\ R.H. Kraichnan,\ J.\ Fluid Mech.\ {\bf 5}, 497 (1959)\\

\noindent  5.\ L. Prandtl,\  Math.\ Mech.\ {\bf 1}, 431 (1925)\\

\noindent  6.\ L.F. Richardson,\ Proc.\ Roy.\ Soc.\ {\bf A10}, 709 (1926)\\

\noindent  7.\ S.A. Orszag and G.S. Patterson,\ ``Numerical
Simulations of Turbulence", in {\it Statistical Models and
Turbulence}, Lecture Notes in Physics {\bf 12}, 127 (1972),
eds.\ M. Rosenblatt and C.W. Van Atta, Springer, Berlin\\

\noindent 8.\ S.A. Orszag,\ in ``Numerical Simulations of the
Taylor-Green vortex", in {\it Computing Methods in Science and
Engineering.}
Part 2. Lecture Notes in Computer Science {\bf 11}, 54 (1974)\\

\noindent 9.\ The traditional estimate is due to L. Landau and
E.M. Lifschitz, Fluid Mechanics, Addison Wesley, Reading, MA,
1959; it was commented upon extensively in the context of spectral
methods by S.A. Orszag,\ ``Statistical theory of turbulence", in
{\it Fluid Dynamics}, Les Houches, 2237 (1973), eds.\
R. Balian and J.L. Peube, Gordon and Breach, New York\\

\noindent 10.\ T. Gotoh, D. Fukayama and T. Nakano, Phys.\ Fluids
{\bf 14}, 1065 (2002); Y. Kaneda, T. Ishihara, M. Yokokawa, K.
Itakura and A. Uno, Phys.\ Fluids {\bf 15}, L21 (2003)\\

\noindent 11.\ P.K. Yeung, D.A. Donzis, Phys.\ Fluids {\bf xx}, Lxx (2005)\\

\noindent 12.\ V. Yakhot, Phys.\ Rev.\ E {\bf 63}, 026307 (2001)\\

\noindent 13.\ R.J. Hill, J.\ Fluid Mech.\ {\bf 368}, 317 (2002)\\

\noindent 14.\ U. Frisch, {\it Turbulence. Legacy of A.N.
Kolmogorov}, Cambridge University Press, England (1995); K.R.
Sreenivasan and R.A. Antonia, Annu.\ Rev.\ Fluid Mech. {\bf 29},
435 (1997)\\

\noindent 15.\ S.Y. Chen, B. Dhruva, S. Kurien
and M.A. Taylor, J.\ Fluid.\ Mech. {\bf 533}, 183, 2005\\

\noindent 16.\  Professor T. Gotoh kindly tested this relation
using the experimental data published in T. Gotoh and T. Nakano,
J.\ Stat.\ Phys. {\bf 113}, 855 (2003); extensive tests using
other sources of data have since been completed to confirm this
result.\\

\noindent 17.\ V. Yakhot, J.\ Fluid Mech.\ {\bf 495}, 135 (2003)\\

\noindent 18.\ V. Yakhot and K.R. Sreenivasan, Physica A {\bf 343}, 147-155, 2004\\

\noindent 19.\ G. Paladin and A. Vulpiani, Phys.\ Rev.\ A {\bf
35}, 1971 (1987); G. Paladin and A. Vulpiani, Phys.\ Rep.\ {\bf
156}, 147 (1990)\\

\noindent 20.\ K.R. Sreenivasan and C. Meneveau, Phys.\ Rev.\ A
{\bf 38}, 6287 (1988); C. Meneveau and K.R. Sreenivasan, J.\ Fluid Mech.\ {\bf 224}, 429 (1991)\\

\noindent 21. M. Nelkin, Phys.\ Rev.\ A {\bf 42}, 7226 (1990); U.
Frisch and M. Vergassola, Europhys.\ Lett. {\bf 15}, 439 (1901)\\

\noindent 22.\ A.M. Polyakov, Phys.\ Rev.\ E {\bf 52}, 6183 (1995)\\

\noindent 23.\ J. Duchon and R. Robert, Nonlinearity {\bf 13}, 249 (2000)\\

\noindent 24.\ G.L. Eyink, Nonlinearity {\bf 16}, 137 (2003)\\

\noindent 25.\  S. Kurien and K.R. Sreenivasan, Phys.~Rev.~E {\bf 64}, 056302 (2001)\\

\noindent 26. R.A. Antonia, M. Ould-Rouis, F. Anselmet and Y. Zhu,
J.\ Fluid Mech.\ {\bf 332}, 395 (1997)\\

\noindent 27. K.R. Sreenivasan, Flow, Turbulence and Combustion
{\bf 72}, 115 (2004)\\

\noindent 28.\ A.S. Monin and A.M. Yaglom, {\it Statistical Fluid
Mechanics, vol.\ 2}, MIT Press, Cambridge, MA (1975)\\

\noindent 29.\ A.M. Reynolds, N. Mordant, A.M. Crawford and E. Bodenschatz, New J.\ Phys.\ art.\ no.\ 58 (2005) \\

\noindent 30.\ M. Chetkov and G. Falkovich, Phys.\ Rev.\ Lett.\
76, 2706 (1996)\\

\noindent 31. O. Roussel and K. Schneider, Computers and Fluids
{\bf 34}, 817 (2005); O.V. Vasilyev and S. Paolucci, J.\ Comp.\
Phys. {\bf 125}, 498 (1996)\\

\noindent 32. Z.-S. She and E. Leveque, Phys.\ Rev.\ Lett. {\bf
72}, 336 (1994)\\

\noindent 33.\ J.W. Deardorff, J.\ Fluid Mech. {\bf 41}, 453 (1970)\\

\noindent 34.\ J. Smagorinsky, Mon.\ Weather Rev.\ {\bf 91}, 99, (1963)\\

\noindent 35.\ V. Yakhot and S.A. Orszag,  J.\ Sci.\ Comp. {\bf 1}, 1 (1986)\\

\noindent 36.\ M. Germano, J.\ Fluid Mech. {\bf 238}, 325 (1992)\\

\noindent 37.\ H.S. Kang, S. Chester and C. Meneveau, J.\ Fluid Mech. {\bf 480}, 129 (2003)\\

\noindent 38.\ L. Larcheveque, P. Segaut, T-H. Le and P. Comte, J.\ Fluid Mech. {\bf 516}, 265 (2004)\\

\noindent 39.\ C. Lee, K. Yeo and J-Il Choi, Phys.\ Rev.\ Lett. {\bf 92}, 144502-1 (2004)\\

\end{document}